\documentclass[prx,preprint,superscriptaddress,amsmath,longbibliography]{revtex4-1}
\usepackage[english]{babel}
\usepackage{graphicx}
\usepackage{color}
\usepackage{float}
\usepackage{graphicx}
\usepackage{amsmath}
\usepackage{amssymb}
\usepackage{gensymb}
\usepackage{multirow}
\usepackage{array}
\usepackage{graphicx}
\usepackage{dcolumn}
\usepackage{bm}
\usepackage{hyperref}
\usepackage{color}
\usepackage{comment}
\usepackage{braket}
\usepackage[version=4]{mhchem}
\usepackage{lineno}

\begin{document}

\title{Multiplexed quantum repeaters \\based on single-photon interference with mild stabilization}

\author{Daisuke Yoshida}
\affiliation{Yokohama National University, 79-5, Tokiwadai, Hodogaya, Yokohama, 240-8501, Japan}
\affiliation{LQUOM Inc., 79-5, Tokiwadai, Hodogaya, Yokohama, 240-8501, Japan}

\author{Tomoyuki Horikiri}
\affiliation{Yokohama National University, 79-5, Tokiwadai, Hodogaya, Yokohama, 240-8501, Japan}
\affiliation{LQUOM Inc., 79-5, Tokiwadai, Hodogaya, Yokohama, 240-8501, Japan}

\date{\today}

\begin{abstract}
{\bf
Quantum repeaters are pivotal in the physical layer of the quantum internet. 
For its development, quantum repeaters capable of efficient entanglement distribution are necessary. 
Quantum repeater schemes based on single-photon interference are promising due to their potential efficiency. 
However, schemes involving first-order interference with photon sources at distant nodes require stringent phase stability in the components,
which poses challenges for long-distance implementation.
In this paper, we present a quantum repeater scheme that leverages single-photon interference with reduced difficulty of phase stabilization. 
Additionally, under specific conditions, we demonstrate that our scheme achieves a higher entanglement distribution rate between end nodes compared to existing schemes.
This approach, implementable using only feasible technologies including multimode quantum memories and two-photon sources, offers high entanglement distribution rates and mild phase stabilization, leading to the development of multimode quantum repeaters.
}
\end{abstract}

\maketitle

\section*{Introduction}
The quantum internet \cite{Kimble2008,Wehner2018,Awschalom2021}, a global network for distributing entanglement, is anticipated to facilitate technological advancements such as distributed quantum computation \cite{Cirac1999,Yimsiriwattana2004,Jiang2007,VanMeter2016},quantum sensor networks~\cite{Jozsa2000,Gottesman2012,Komar2014, Guo2020}, and end-to-end quantum key distribution \cite{Bennett1992,Xu2020,Pirandola2020}. 
For the globalization of entanglement distribution, it is considered essential to have quantum repeaters that expand quantum entanglement from between elementary links to between end nodes through entanglement swapping~\cite{Briegel1998, Duan2001, Sangouard2011}.

One of the promising configurations for quantum repeaters involves each node having an absorptive memory and a two-photon source~\cite{simon2007}. 
Quantum repeaters using a built-in type of quantum memory in which the photon source and memory are integrated are also possible \cite{Duan2001, Chou2005,Delteil2016, Stockill2017, Humphreys2018, Stephenson2020, Pompili2021, Liu2023multinode}. 
Most quantum memories do not emit photons in the telecommunication wavelength band.
On the other hand, in a configuration using absorptive memory, a nondegenerate two-photon source can be used to select the telecommunication wavelength band for long-distance transmission~\cite{Fekete2013,Aizawa2023,LagoRivera2023,Rakonjac2023}.
In addition, time-division multiplexing, which is advantageous for high communication rates, has recently been demonstrated by two experimental groups \cite{Lago-Rivera2021, Liu2021}.
Although the configuration is similar in these two demonstrations, the entanglement generation schemes are different.
Ref.~\cite{Lago-Rivera2021} demonstrated a single-photon interference-based single-excitation entanglement generation scheme (SS), while Ref.~\cite{Liu2021} demonstrated a two-photon interference-based two-excitation entanglement generation scheme (TT).
SS is more efficient than TT in scenarios with low photon pair generation rates or high optical transmission losses, such as long elementary link distances \cite{Wu2020}.
This is because TT's entanglement heralding relies on two-photon coincidence counting at the central station, whereas SS uses single-photon detection. 
However, SS requires precise control or monitoring of the paths of the signal at sub-wavelength scales as well as the phase of the lasers pumping the two-photon sources, presenting significant challenges in phase control.

In this paper, we propose a single-photon interference-based two-excitation entanglement generation scheme (ST).
This scheme significantly reduces the phase requirement by over two orders of magnitude compared to SS~\cite{Duan2001, Chou2005,Delteil2016, Stockill2017, Humphreys2018,Pompili2021,Lago-Rivera2021, Liu2023multinode}, while preserving the high efficiency of entanglement generation. 
We demonstrate that ST can outperform the SS in entanglement generation rates under specific conditions and discuss potential setups for implementing ST.

Our approach is inspired by the mode-pairing quantum key distribution \cite{Zeng2022, Xie2022}, and a frequency multiplexed quantum repeater~\cite{Sinclair2014}.

\section*{Results}
\subsection*{Components of the proposed scheme}
Fig.~\ref{STdia}a presents a schematic diagram of the entanglement generation process in an elementary link using the ST.
Prior to detailing the procedure, we outline the function of each component as follows.

\subsubsection*{A two-photon source}
Fig.~\ref{STdia}c illustrates an overview of the two-photon source (TPS) utilized in the ST.
The TPS generates photon pairs across $M$ frequency modes and $N$ temporal modes in each trial, with a frequency mode interval of $\mathit{\Delta_{\mathrm{f}}}$ and a temporal mode interval of $\mathit{\Delta_{\mathrm{t}}}$.
The probability of photon pair emission in each mode is $p_{\mathrm{tps}}$.
One photon of each pair is designated as the signal photon, transmitted to the adjacent multimode quantum memory (MQM), and the other as the idler photon, sent to the central Bell state analyzer (CBSA) via a long-distance transmission line.
These signals and idlers are phase-correlated, as is typical in spontaneous parametric down-conversion (SPDC) processes. 
In SPDC, the sum of the phases of signal and idler photons is determined by the pump laser's phase~\cite{Ou1989, Heuer2014}.
We assume that the sum of the phases for signal and idler photons remains constant across all modes within the same trial at each node in this scheme.
The sum of phases at node $Z$ is denoted as $\mathit{\theta_Z}$.
The frequencies of idler and signal photons in the $m$-th frequency mode at node $Z$ are represented as $f_{Z\mathrm{i},m}$ and $f_{Z\mathrm{s},m}$, respectively.
In this scheme, idler photons from adjacent nodes are assumed to have identical frequencies in each mode, which is required for interference at the CBSA, i.e. $f_{\mathrm{i},m}=f_{A\mathrm{i},m}=f_{B\mathrm{i},m}$.
Additionally, we assume that the sum of the frequencies of idler and signal photons remains constant, i.e., $f_{\mathrm{i},m}+f_{Z\mathrm{s},m}=f_{\mathrm{i},m'}+f_{Z\mathrm{s},m'}$.
This assumption aligns with the principles of SPDC, where typically, the combined frequencies of these photon pairs equal the frequency of the pump laser.
Although SPDC processes can potentially produce multiple photon pairs, we operate under the assumption that the probability of such occurrences is negligible~\cite{Wu2020, Sinclair2014}.
Thus, the state of photon pairs at node $Z$ in one trial is represented as
$\sum_{n=1}^N \sum_{m=1}^M (\sqrt{1-p_\mathrm{tps}} + 
e^{i\theta_Z}\sqrt{p_\mathrm{tps}}z^\dagger_{\mathrm{s}, mn}z^\dagger_{\mathrm{i}, mn})\ket{0}$,
where $z_{\mathrm{s(i)}, mn}$ signifies the mode for the $m$-th frequency and $n$-th temporal mode signal (idler) photon at node $Z$, and $\ket{0}$ is the vacuum state.

\subsubsection*{A multimode quantum memory}
\begin{figure}[h]
    \centering
    \includegraphics[width=0.95\textwidth]{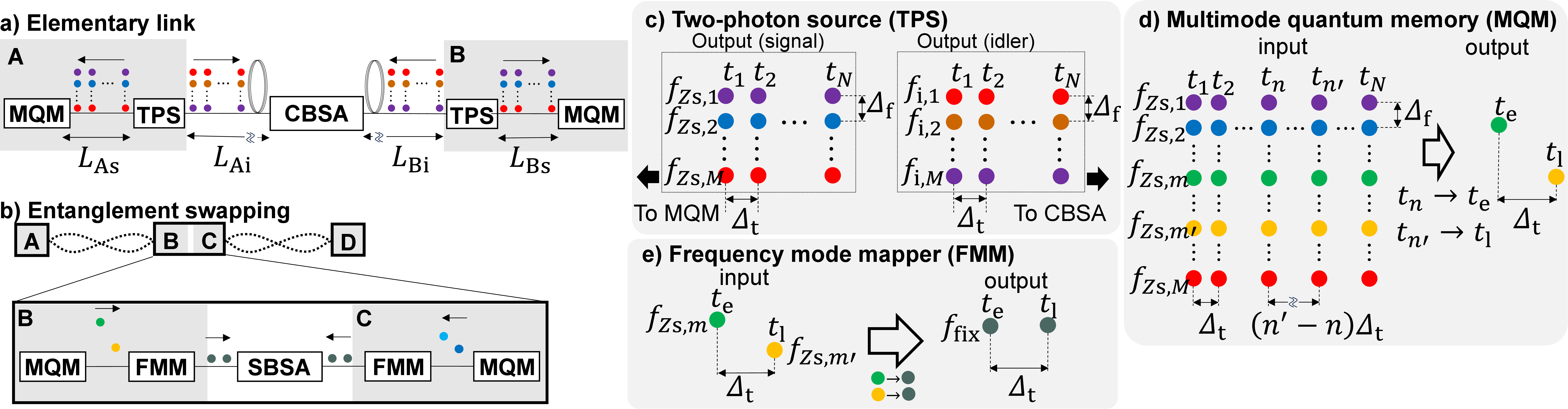}
    \caption{
    {\bf Schematic diagram of the ST.}
    {\bf a}, Schematic diagram of elementary link.
    One photon of each pair emitted by the two-photon source (TPS) is transmitted to a remote station and inputted into the central Bell state analyzer (CBSA). The other photon is inputted into a multimode quantum memory (MQM). 
    $L_\mathrm{A(B)i}$ denotes the optical path length from Node A(B)'s TPS to the CBSA. 
    $L_\mathrm{A(B)s}$ represents the optical path length from Node A(B)'s TPS to MQM.
    {\bf b}, Overview of entanglement swapping in the ST. Photons are regenerated from the MQM to coincide in timing with photons regenerated from adjacent MQMs. 
     Additionally, frequency mode mappers (FMMs) are employed to achieve indistinguishability in the frequency domain of photons inputted into the Swapping Bell State Analyzer (SBSA).
    {\bf c}, Schematic representation of the TPS.
    {\bf d}, Schematic diagram of the MQM.
    {\bf e}, Overview of the FMM process.
    }
    \label{STdia}
\end{figure}

Fig. \ref{STdia}d provides an overview of the MQM. 
We assume that the MQM at each node can store signal photons in $M$ frequency modes and $N$ temporal modes, generated by an adjacent TPS. 
The MQM is also assumed to preserve the phase of each input photon across all modes and retrieve them at any desired time. 
Essentially, the MQM can transform input photons into any temporal modes. 
Within the ST, this capability enables the conversion of photons from two different modes into temporal modes $t_\mathrm{e}$ and $t_\mathrm{l}$, respectively, forming a time-bin qubit with a time interval $\mathit{\Delta_{\mathrm{t}}}$.
As depicted in Fig. \ref{STdia}d, the MQM, for instance, converts photons in temporal modes $n$ and $n'$ into $t_\mathrm{e}$ and $t_\mathrm{l}$. We also assume that the MQM's efficiency, denoted as $\eta_{\mathrm{QM}}$, is uniform across all modes.

\subsubsection*{A frequency mode mapper}
Fig \ref{STdia}e shows an overview of a frequency mode mapper (FMM).
We assume that the FMM can convert the frequency of an incoming photon to a pre-determined frequency, denoted as $f_{\mathrm{fix}}$. 
As illustrated in Fig \ref{STdia}e, the FMM, for example, modulates photons in frequency modes $f_{Z\mathrm{s},m}$ and $f_{Z\mathrm{s},m'}$ to $f_{\mathrm{fix}}$. 
We also assume that the modulation efficiency of the FMM, $\eta_{\mathrm{FM}}$, is consistent across all modes.

\subsubsection*{A Bell state analyzer}
In this scheme, two types of BSAs are utilized: a CBSA, located between repeater nodes, and a swapping BSA (SBSA) for entanglement swapping.
The CBSA transforms two input modes, $a$ and $b$, into two output modes, $d_+$ and $d_-$, with $d_\pm = (a \pm b)/\sqrt{2}$ \cite{Duan2001}.
Each output mode is equipped with a photon detector capable of identifying the temporal and frequency modes of detected photons.
In the ST, detecting a single photon in either $d_+$ or $d_-$ mode in each trial signifies successful entanglement generation between elementary links.
The SBSA is a type-2 fusion gate-based BSA designed for time-bin qubits with a time interval $\mathit{\Delta_{\mathrm{t}}}$ \cite{Soudagar2007}.

\subsection*{Description of the proposed scheme}
\subsubsection*{Elementary link}
In this section, we describe the entanglement generation procedure of the ST between nodes A and B, forming an elementary link. 
The generation of each photon pair is timed so that photons in the same temporal mode arrive simultaneously at the CBSA.
The phase acquired by the idler (signal) photon of the $m$-th frequency mode and $n$-th temporal mode at node $Z$, on its path from the TPS to the CBSA (MQM), is $2\pi f_{\mathrm{i(\mathit{Z}s)},m} L_{Z\mathrm{i}(Z\mathrm{s}),n}/c$, where $L_{Z\mathrm{i}(Z\mathrm{s}),n}$ represents the optical path length and $c$ is the speed of light in the optical path.
We assume that during $N\mathit{\Delta_{\mathrm{t}}}$, the fluctuations in the optical path lengths are sufficiently small compared to the wavelengths.
Consequently, the optical path length is independent of the temporal mode, i.e.,$L_{Z\mathrm{i}(Z\mathrm{s})}=L_{Z\mathrm{i}(Z\mathrm{s}),n}=L_{Z\mathrm{i}(Z\mathrm{s}),n'}$.
Detection of a photon in the $m$-th frequency mode and $n$-th temporal mode at $d_\pm$ results in the following states between the memories:
\begin{equation}
\ket{\Psi_\pm}_{\mathrm{AB},mn}=
\frac{1}{\sqrt{2}}
(a^\dagger_{\mathrm{s},mn} \pm 
e^{i\theta_{\mathrm{AB}, m}} 
b^\dagger_{\mathrm{s},mn}) \ket{0},
\label{SS}
\end{equation}
where,
\begin{align}
\theta_{\mathrm{AB}, m} = \frac{2\pi}{c}
(
f_{\mathrm{i},m}\mathit{\Delta_{L_\mathrm{i}}} 
+ f_{B\mathrm{s},m} L_{B\mathrm{s}}
- f_{A\mathrm{s},m} L_{A\mathrm{s}}
)
+ \mathit{\Delta_{\theta}},
\label{phaseSS}
\end{align}
with $\mathit{\Delta_{L_\mathrm{i}}} = L_\mathrm{Bi} - L_\mathrm{Ai}$ and $\mathit{\Delta_{\theta}} = \theta_\mathrm{B} - \theta_\mathrm{A}$.
The global phase is omitted for simplicity.
The state described in Eq. \eqref{SS} is identical with the state in the SS. 

Now, let's consider the scenario where one photon in the $m$-th frequency mode and $n$-th temporal mode, and another photon in the $m'$-th frequency mode and $n'$-th temporal mode, are detected in $d_+$.
When extracting only the case where each of the nodes A and B has one excitation from the state $\ket{\Psi_+}_{\mathrm{AB},mn} \otimes \ket{\Psi_+}_{\mathrm{AB},m'n'}$, the resulting state between the memories is given by:
\begin{equation}
\begin{split}
\ket{\Psi_{++}}_{\mathrm{AB},mnm'n'} = 
&\frac{1}{\sqrt{2}}
(a^\dagger_{\mathrm{s},mn} b^\dagger_{\mathrm{s},m'n'}\\&+ 
e^{i\theta_{\mathrm{AB}, mm'}} 
a^\dagger_{\mathrm{s},m'n'} b^\dagger_{\mathrm{s},mn}) \ket{0},
\label{ST1}
\end{split}
\end{equation}
where,
\begin{align}
\begin{split}
\theta_{\mathrm{AB}, mm'} = 
\frac{2\pi(m-m') \mathit{\Delta}_\mathrm{f}}{c}
\big(\mathit{\Delta_{L_\mathrm{i}}}-\mathit{\Delta_{L_\mathrm{s}}}\big),
\label{phaseST}
\end{split}
\end{align}
with $\mathit{\Delta_{L_\mathrm{s}}} = L_\mathrm{Bs} - L_\mathrm{As}$.
As in Eq. \eqref{SS}, the global phase is omitted in Eq. \eqref{ST1}.
The state in Eq. \eqref{ST1} represents the two-excitation entanglement created between the memories of elementary links in the ST.

Comparing the relative phase $\theta_{\mathrm{AB}, m}$ (Eq. \eqref{phaseSS}) in the SS with $\theta_{\mathrm{AB}, mm'}$ (Eq. \eqref{phaseST}) in the ST highlights two key points.
Firstly, the impact of optical path fluctuations on the phase is significantly smaller in the ST.
In SS, the phase term involves the product of each photon's frequency and the optical path, whereas in ST, it involves the product of the frequency mode spacing and the optical path length difference. 
Given that photon frequencies are typically around $\sim10^{14} ~\mathrm{Hz}$ and frequency mode spacing can be much smaller, e.g., $\sim10^{10} ~\mathrm{Hz}$, the optical path stability required for equivalent entanglement fidelity in ST is reduced by approximately four orders of magnitude compared to SS.
Secondly, in ST, the sum of the photon pair initial phases does not influence the entangled state's phase.
In SS, $\theta_{\mathrm{AB}, m}$ can vary each trial as $\mathit{\Delta_{\mathrm{\theta}}}$ affects the phase, but in ST, $\mathit{\Delta_{\mathrm{\theta}}}$ is a global phase and does not impact $\theta_{\mathrm{AB}, mm'}$.
When SPDC is utilized as the TPS, where the photon pair's phases are determined by the pump laser's phase, the ST only necessitates a pump laser with a coherence time longer than $N\mathit{\Delta_{\mathrm{t}}}$.
This means phase synchronization or monitoring the phase difference between remote lasers is not required in ST.

In Eq. \eqref{ST1}, the combination of temporal and frequency modes can vary with each trial, posing challenges for entanglement swapping and other applications.
To address this, the modes of retrieved photons are adjusted to a predetermined mode using MQMs and FMMs.
By converting the modes of $\ket{\Psi_{++}}_{\mathrm{AB},mm'}$ and applying phase modulation to achieve $\theta_{\mathrm{AB}, mm'} =0$, we obtain the following time-bin entangled state $\ket{\Psi_{++}}_{\mathrm{AB}}$:
\begin{equation}
\ket{\Psi_{++}}_\mathrm{AB} = 
\frac{1}{\sqrt{2}}
(a^\dagger_{\mathrm{s,e}} b^\dagger_{\mathrm{s,l}}+ 
a^\dagger_{\mathrm{s,l}} b^\dagger_{\mathrm{s,e}}) \ket{0}.
\label{ST2}
\end{equation}

\subsubsection*{\label{level3-6}Entanglement swapping}
This section describes the entanglement swapping process between the heralded states $\ket{\Psi_{++}}_{\mathrm{AB}}$ and $\ket{\Psi_{++}}_{\mathrm{CD}}$, as illustrated in Fig \ref{STdia}b. 
Previously, we focused on cases where two memories in an elementary link are each excited once. 
Now, we extend our analysis to include scenarios where one memory is doubly excited, resulting in the combined state 
$\frac{1}{4}
(a^\dagger_{\mathrm{s,e}} b^\dagger_{\mathrm{s,l}}
+a^\dagger_{\mathrm{s,l}} b^\dagger_{\mathrm{s,e}}
+a^\dagger_{\mathrm{s,e}} a^\dagger_{\mathrm{s,l}}
+b^\dagger_{\mathrm{s,e}} b^\dagger_{\mathrm{s,l}})\otimes
(c^\dagger_{\mathrm{s,e}} d^\dagger_{\mathrm{s,l}}
+c^\dagger_{\mathrm{s,l}} d^\dagger_{\mathrm{s,e}}
+c^\dagger_{\mathrm{s,e}} c^\dagger_{\mathrm{s,l}}
+d^\dagger_{\mathrm{s,e}} d^\dagger_{\mathrm{s,l}}) \ket{0}.
$
To avoid successful entanglement swapping when one memory is doubly excited, we use a type-2 fusion gate-based Bell measurement in the SBSA~\cite{Chen2007, Soudagar2007}.
Ideally, with perfect device efficiencies, successful Bell measurement events happen only when each node has exactly one excited memory. 
However, in practice, due to losses in each device, the resultant state $\rho_{\rm{AD}}$ between nodes A and D after a successful Bell measurement is expressed as:
\begin{equation}
    \rho_{\rm{AD}} = \alpha_{11} \rho_{11} + \alpha_{1} \rho_{1} + \alpha_{0} \rho_{0},
\label{rho_AD}
\end{equation} 
where $\rho_{11}$ is the state with one excitation in both memories, $\rho_{1}$ represents the state with a single excitation across the two memories, and $\rho_{0}$ is the vacuum state. The coefficients $\alpha_{11}$, $\alpha_{1}$, and $\alpha_{0}$ correspond to the probabilities of these states. (For more details, please refer to Supplementary Information Note 1).

\subsection*{Performance comparison}
Next, we compare the coincidence rates of photons emitted from the memory of end nodes in the SS, ST, and TT. 
For this comparison, we consider a symmetric setup where the distance $L/2$ from each node to the CBSM is identical.
Additionally, we assume that the coherence time is sufficiently long to ensure that the efficiency of the MQM is not affected by storage time.
We also presume that the dark count of the photon detectors is negligible.
In practice, detectors with counts lower than Hz have been achieved~\cite{Liu2023}.
The detection efficiency of the photon detectors is denoted as $\eta_\mathrm{det}$. 
To ensure a fair comparison, we suppose that the SS and TT have the same multiplicity as the ST and that mode conversion is feasible.

\begin{figure}[h]
    \centering
    \includegraphics[width=0.5\textwidth]{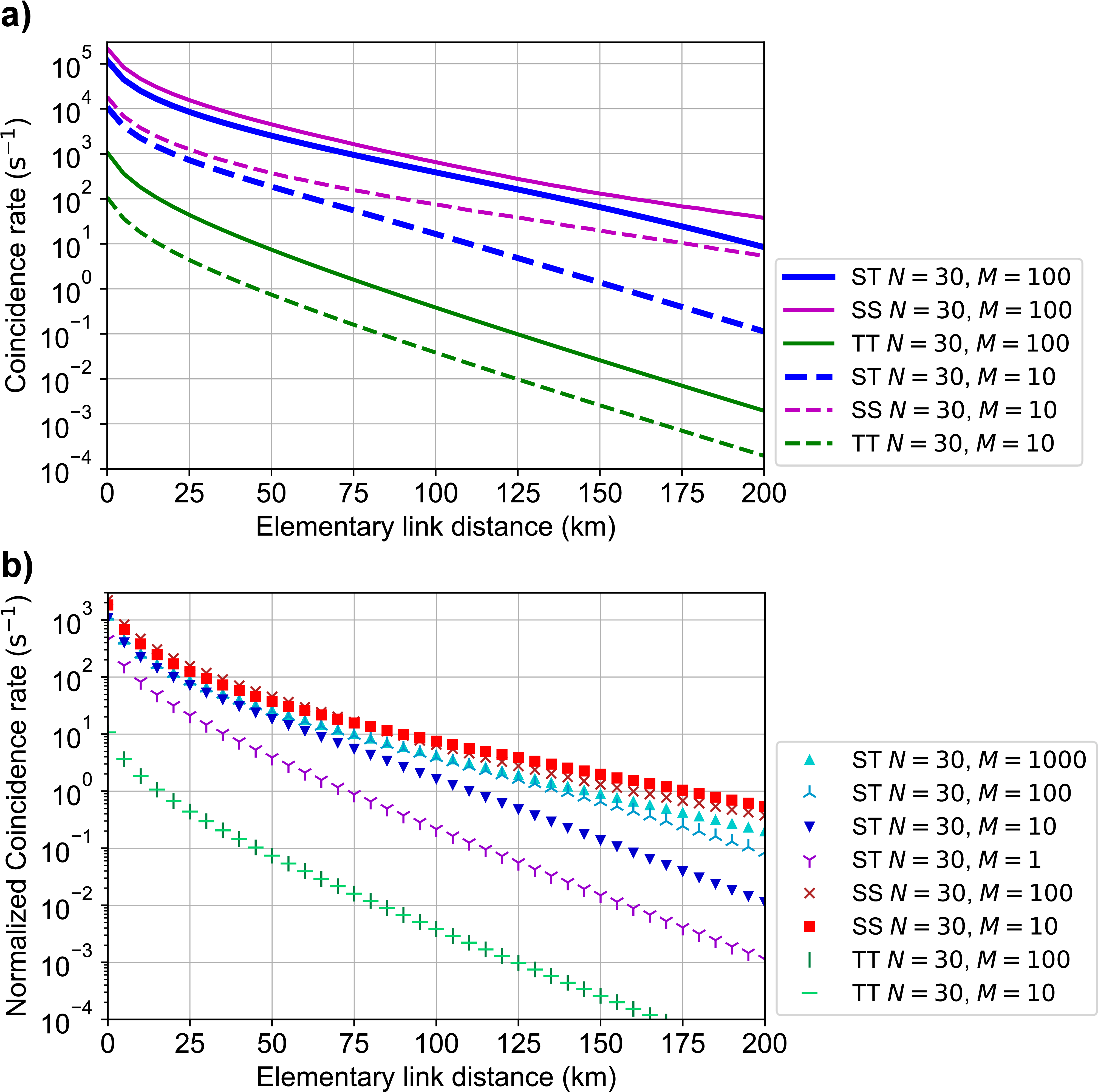}
    \caption{
    {\bf Comparison of each scheme in elementary links.}
    {\bf a}, Computed coincidence rates for each scheme within the elementary link. 
    The horizontal axis represents the inter-node distance. 
    Solid lines depict results with a frequency mode number $M=100$, while dashed lines represent $M=10$.
    For both, the temporal mode number is set at $N=10$.
    The dotted line considers the original DLCZ scheme, assuming no multiplexing for both time and frequency modes; hence, no FMM is performed. 
    Other parameters include detector efficiency $\eta_{\mathrm{det}} = 0.9$, quantum memory storage efficiency $\eta_{\mathrm{QM}} = 0.5$, temporal mode interval $\mathit{\Delta_{\mathrm{t}}} = 620 \mathrm{ns}$, and an optical path loss of 0.2dB/km.
    {\bf b}, Results from {\bf a} are normalized by the number of frequency modes.
    Reference fiber is the nearest neighbor of Signal fiber.
    }
    \label{fig:ele}
\end{figure}

We first analyze the coincidence rate in the elementary link. 
Let $K_X^{(0)}$ represents the number of heralding events for entanglement in each scheme per trial, where $X = \mathrm{SS, ST, TT}$. 
In the SS, $K_\mathrm{SS}^{(0)}$ is the number of modes where a single photon detection event occurs in CBSM. 
Conversely, in the TT, $K_\mathrm{TT}^{(0)}$ equals the number of modes with a specific two-photon detection event in CBSM~\cite{Chen2007, Xu2017}.
For the ST, $K_\mathrm{ST}^{(0)} = \lceil K_\mathrm{SS}^{(0)}/2 \rceil$, as it is established by pairing two single-excitation entanglements.
Consequently, the expected values of $K_\mathrm{SS}^{(0)}$, $K_\mathrm{TT}^{(0)}$, and $K_\mathrm{ST}^{(0)}$ are calculated as follows:
\begin{align}
    E(K_\mathrm{SS}^{(0)}) &= 2NMp_{\mathrm{tps}} \eta_{\mathrm{det}} \eta_{\mathrm{att}}\left(\frac{L}{2}\right), \label{eq:KSS0}
\end{align}
\begin{align}
    E(K_\mathrm{TT}^{(0)}) &= \frac{1}{2} NM \left( p_{\mathrm{tps}} \eta_{\mathrm{det}} \eta_{\mathrm{att}}\left(\frac{L}{2}\right) \right)^2 \label{eq:KTT0}, 
\end{align}
\begin{equation}
\begin{split}
    E(K_\mathrm{ST}^{(0)}) &= \left(\sum_{k=0}^{\frac{NM}{2}-1} k \left(P_\mathrm{h} (2k) + P_\mathrm{h} (2k+1)\right)\right) \\
    &+ \frac{NM}{2} P_\mathrm{h} (NM) \label{eq:KST0},
\end{split}
\end{equation}
where $\eta_{\mathrm{att}}\left(\frac{L}{2}\right)$ represents the transmittance for a path length of $L/2$, and $NM$ is assumed to be an even number.
$P_\mathrm{h} (k)$ is defined as ${}_{NM} \mathrm{C}_k (2p_{\mathrm{tps}} \eta_{\mathrm{det}} \eta_{\mathrm{att}}(\frac{L}{2}))^k (1-2p_{\mathrm{tps}} \eta_{\mathrm{det}} \eta_{\mathrm{att}}(\frac{L}{2}))^{NM-k}$.
From Eq. \eqref{eq:KSS0} and \eqref{eq:KTT0}, it's evident that $E(K_\mathrm{SS}^{(0)})$ is linearly affected by various efficiencies, whereas $E(K_\mathrm{TT}^{(0)})$ is impacted quadratically.
Fig. \ref{fig:ele}a displays the calculated two-photon coincidence rate for photons retrieved from both memories in the elementary link. 
The derivation methods are detailed in the Methods section. 
The ST's rate decays at a slope similar to the SS in low transmission loss regions, and closer to the TT in high loss regions. 
Notably, there are distances where the SS and ST rates are comparable. 
Fig. \ref{fig:ele}b then presents the coincidence rate normalized by the frequency mode number. 
In the ST, the distance at which the rate's slope changes depends on the frequency mode number. 
Interestingly, even with a relatively small number of modes, such as $M = 1$, the normalized rate in the ST exceeds that of the TT by more than an order of magnitude.

\begin{figure}[h]
    \centering
    \includegraphics[width=0.5\textwidth]{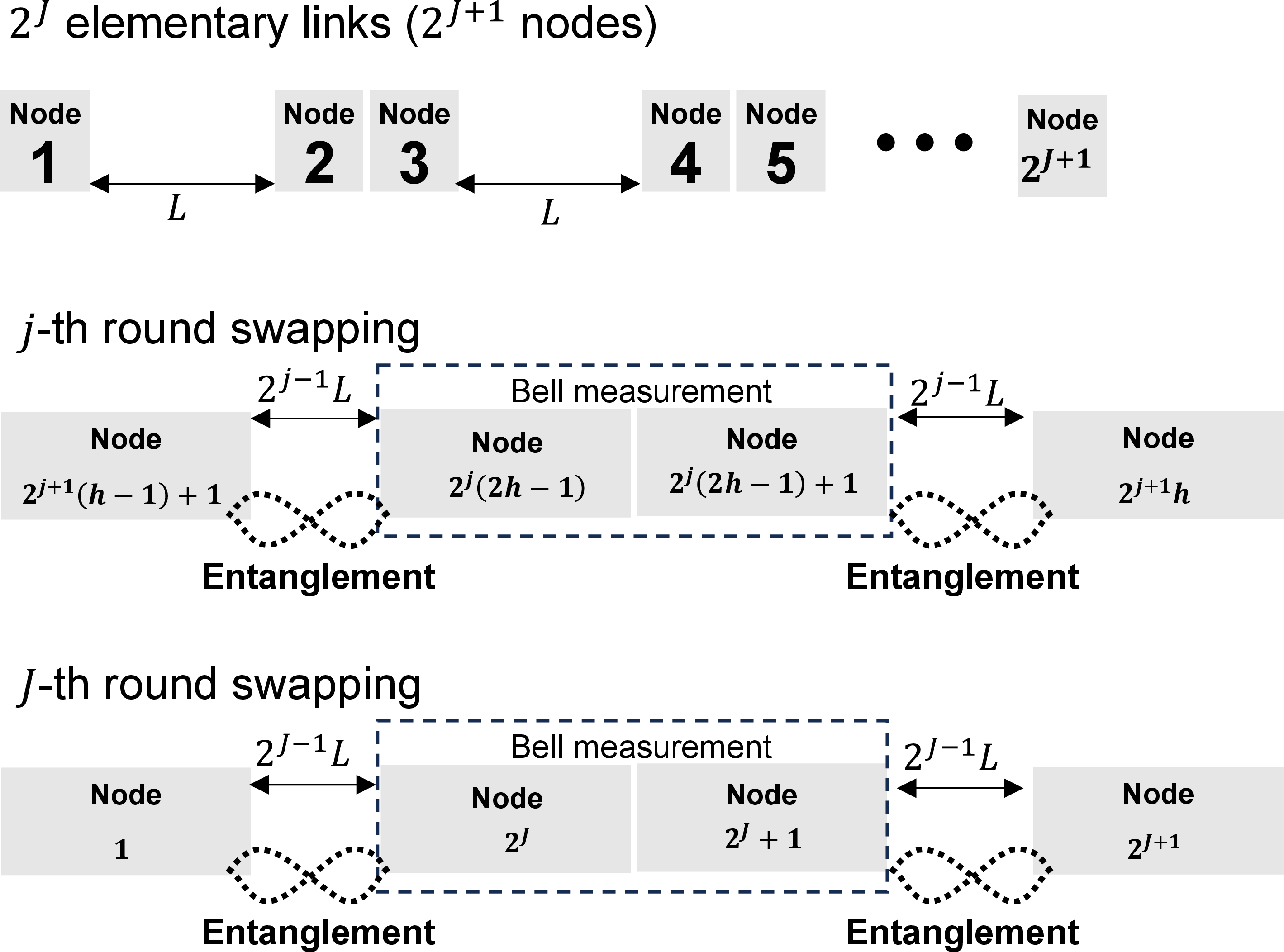}
    \caption{
    {\bf Diagram of a transmission path composed of $2^J$ elementary links.}
    In the $j$ th round of swapping, Bell measurements are performed at nodes $2^{j}(2h-1)$ and $2^{j}(2h-1)+1$, attempting to generate entanglement between these $2^{j+1}(h-1)+1$ and $2^{j+1}h$ nodes.
    }
    \label{fig:swap}
\end{figure}

Next, we analyze the coincidence rate following entanglement swapping. 
Let's assume that the communication channel consists of $2^J$ elementary links ($J \geq 1$) and involves $J$ rounds of swapping, as shown in Fig.~\ref{fig:swap}.
In the $j$-th round of swapping, Bell measurements are conducted on nodes $2^j(2h-1)$ and $2^j(2h-1)+1$ ($h \leq 2^{J-j}$).
The total time $T^{(J)}_{\mathrm{tot}, X}$, starting from the beginning of the trial and leading up to the point of achieving a single coincidence event from photons retrieved at the end node’s memory, is expressed for $X = $ST or TT as follows:
\begin{align}
    T^{(J)}_{\mathrm{tot}, X} = 
    \sum_{j = 0}^{J-1} \left\{
    t^{(j)}
    \frac{1}{p_X^{(J)}p_{\mathrm{ps,} X}^{(J)}}
    \prod_{h = j}^{J-1}\left(
    n_\mathrm{ex}(p_X^{(h)})
    \right)
    \right\}.
    \label{eq:RSTTT}
\end{align}
For $X = $SS, where coincidence is attempted at the end node using single excitation entanglement in two different modes, it is represented as:
\begin{align}
    T^{(J)}_{\mathrm{tot}, X} = 
    \sum_{j = 0}^{J} \left\{
    t^{(j)}
    \frac{1}{p_{\mathrm{ps,} X}^{(J)}}
    \prod_{h = j}^{J}\left(
    n_\mathrm{ex}(p_X^{(h)})
    \right)
    \right\}.
    \label{eq:RSS}
\end{align}
Here, $t^{(j)}$  is the time required to inform the necessary node of the results from the $j$-th round of entanglement swapping ($2^{j-1}L/c$ for $j \geq 1$) or the time for measurement results from CBSM to reach the adjacent node after initiating entanglement distribution on the elementary link ($L/c + N\mathit{\Delta}_\mathrm{t}$ for $j = 0$).
$p_X^{(h)}$ is the probability of successful entanglement swapping in the $h$-th round for $h \geq 1$, and the probability of heralding at least one entanglement on the elementary link for $h = 0$.
$n_\mathrm{ex}(p)$ is the expected number of trials for two independent events with success probability $p$ to both succeed at least once, given by $n_\mathrm{ex}(p)=\frac{3-2p}{(2-p)p}$.
$p_{\mathrm{ps,} X}^{(J)}$ is the probability of achieving coincidence at the end node. 
It is also assumed that even if multiple entanglements are heralded between memories, only one mode can be swapped.
Details on each probability calculation are provided in Supplementary Information Note 1.

Fig \ref{Figjlink} depicts the coincidence rate $R^{(J)}_\mathrm{X}=1/T^{(J)}_{\mathrm{tot}, X}$ for $J = 1, 2, 3, 4, 5$, where the number of links has been optimized for each distance between end nodes to maximize the rate. 
The parameters used in these calculations are listed in Fig.~\ref{Figjlink}'s caption. 
In practical scenarios (represented by dashed lines), the ST's rate is 1-2 orders of magnitude lower than the SS.
This large difference mainly stems from the different swapping requirements between two-excitation and single-excitation entanglement: single-excitation entanglement involves detecting a single photon, while two-excitation entanglement requires coincidence detection \cite{Wu2020}.
In the optimistic scenario (solid lines), the difference between ST and SS is smaller.
Interestingly, some end-node distances show the ST outperforming SS.
This can be understood by examining the state $\rho_{X}^{(J)}$ between end-node memories after a successful $J$-th swapping operation, represented as follows~\cite{Wu2020, Chen2007}:
\begin{align}
    \rho_{\rm{SS}}^{(J)} = \alpha^{(J)} \rho_{1} + (1 - \alpha^{(J)}) \rho_{0}, \label{eq:rhoSS}
\end{align}
\begin{align}
    \rho_{\rm{ST}}^{(J)} = \alpha_{11}^{(J)} \rho_{11} + \alpha_{1}^{(J)} \rho_{1} + \alpha_{0}^{(J)} \rho_{0}, \label{eq:rhoST}
\end{align}
\begin{align}
    \rho_{\rm{TT}}^{(J)} = \rho_{11}. \label{eq:rhoTT}
\end{align}
In SS, $\rho_1$ is the desired state, while $\rho_0$ is not. 
$\alpha^{(J)}$ indicates the probability of $\rho_\mathrm{SS}$ being $\rho_1$. 
For ST, $\rho_{11}$ is desired, with $\rho_1$ and $\rho_0$ as undesired states. 
The coefficients for each state, influencing the likelihood of successful coincidence events, are described as:
\begin{align}
    \alpha^{(J)} = \frac{\alpha^{(J-1)}}{2-\alpha^{(J-1)}\eta}, \label{eq:alpha}
\end{align}
\begin{align}
    \alpha_{11}^{(J)} = \frac{1}{\left(1 + \frac{\alpha_{1}^{(1)}}{2\alpha_{11}^{(1)}}\right)^2}, \label{eq:alpha11}
\end{align}
\begin{align}
    \alpha_{1}^{(J)} = \frac{\alpha_{1}^{(1)}}{\alpha_{11}^{(1)}\left(1 + \frac{\alpha_{1}^{(1)}}{2\alpha_{11}^{(1)}}\right)^2}, \label{eq:alpha1}
\end{align}
\begin{align}
    \alpha_{0}^{(J)} = \left(\frac{\alpha_{1}^{(1)}}{2\alpha_{11}^{(1)} + \alpha_{1}^{(1)}}\right)^2. \label{eq:alpha0}
\end{align}
Here, $\eta$ is defined as $\eta = \eta_\mathrm{QM}\eta_\mathrm{FM}\eta_\mathrm{det}$.
Given that $2 - \alpha^{(J)}\eta > 1$, $\alpha^{(J)} < \alpha^{(J-1)}$ as per Equation \eqref{eq:alpha}, suggesting a decrease in the coefficient for the desired state $\rho_1$ in SS as swapping operations increase. 
Conversely, in the ST, according to Equations \eqref{eq:alpha11} to \eqref{eq:alpha0}, the coefficients are constant regardless of the number of swapping operations. 
This constancy might explain why the rate of the ST surpasses that of the SS in certain regions.
For cases where $J\geq2$, the expressions in Eqs. \eqref{eq:alpha11} to \eqref{eq:alpha0} are applicable.
For the specific case of $J = 1$, please refer to the expressions provided in Supplementary Information Note 1.

\section*{Discussion}
We consider a potential configuration for ST, possibly involving the use of an Atomic Frequency Comb (AFC)~\cite{Afzelius2009} created within the inhomogeneous broadening of a \ce{Pr^{3+}}:\ce{Y2SiO5} crystal (Pr:YSO) as the MQM, cavity-enhanced SPDC (cSPDC) as the TPS~\cite{Goto2003}, and a phase modulator as the FMM~\cite{Johnson1988}.
The AFC offers extensive multiplexing capabilities.
Particularly, the AFC in Pr:YSO, with a frequency mode interval of 100 MHz and a temporal mode interval of 620 ns, is estimated to store about 100 frequency modes and 32 temporal modes with an efficiency exceeding 50\%~\cite{Ortu2022}.
Already demonstrated is the on-demand storage of 30 temporal modes~\cite{Ortu2022}.
Additionally, storing multiple frequency mode photons emitted from cSPDC has been achieved~\cite{Seri2019, Ito2023}.
Regarding the FMM, employing a phase modulator for serrodyne modulation enables frequency modulation without creating surplus modes~\cite{Johnson1988}.
The modulation bandwidth of the FMM should be about $M\mathit{\Delta_{\mathrm{f}}}$, which is around 10 GHz when using a Pr:YSO-based AFC as MQM. 
With high-bandwidth signal generators and phase modulators, serrodyne modulation of several GHz or more is feasible~\cite{Saglamyurek2014}.
For sub-ns pulses, modulation of 20 GHz has been achieved with nearly 100\% efficiency~\cite{Puigibert2017}.
 
Now, let’s consider the stabilization of the relative phase $\theta_{\mathrm{AB}, mm'}$ (Eq. \eqref{phaseST}) in the entangled state produced by ST.
Assuming that the phase fluctuation follows a normal distribution with a standard deviation $\sigma$, the fidelity F can be expressed as $(1+e^{-\frac{\sigma^2}{2}})/2$ \cite{Minar2008}.
For a fidelity of 0.99, $\sigma$ is approximately 0.2 rad, which serves as our reference value.
For $(m-m')\mathit{\Delta_{\mathrm{f}}}$ equal to 10 GHz, a displacement of about 1 mm in $\mathit{\Delta_{L\mathrm{i}}}-\mathit{\Delta_{L\mathrm{s}}}$ results in a phase shift of roughly 0.2 rad.
Here, $\mathit{\Delta_{L\mathrm{s}}}$ denotes the difference in light paths within a node, which are relatively short and more easily controlled for environmental factors such as temperature.
Conversely, $\mathit{\Delta_{L\mathrm{i}}}$ represent the difference in longer light paths, which could be field-deployed fiber, and are thus more susceptible to uncontrolled environmental conditions.
Therefore, it is assumed that the dominant factor influencing $\sigma$ is the fluctuation in $\mathit{\Delta_{L\mathrm{i}}}$, with the aim being to stabilize this fluctuation to within approximately 1 mm.
This stabilization could potentially be achieved by transmitting a reference pulse from each node to the CBSM, measuring the time difference in arrival with a high-speed detector, and subsequently providing feedback to a piezoelectric actuator or a stage.
Moreover, employing a wavelength different from that of the idler photons for the reference light in this stabilization process could allow for nearly 100\% duty cycle in entanglement distribution.

Finally, we emphasize the significance of reducing the requirements for optical path length stability and phase stability of the pump lasers in TPSs, which is a key aspect of the ST.
Stringent requirements for optical path stability lead to frequent need for monitoring or stabilizing the phase difference. 
This necessitates multiple interventions before an entangled state is heralded between end nodes, potentially complicating both operational procedures and the design of the optical system.
For instance, in the process of entanglement swapping between adjacent quantum memories, fluctuations in the optical path on the order of sub-$\mathrm{\mu}$m can occur while waiting for entanglement heralding in both memories.
In the SS, simultaneously monitoring or stabilizing the interferometer's phase difference and storing photons in quantum memory is not straightforward.
However, by sufficiently relaxing these requirements, such challenges can be mitigated.
Additionally, even with a stable optical path, when using SPDC for photon pairs, the SS would require monitoring or locking the phase difference of pump lasers between distant nodes.
This might require a more complex system involving optical frequency combs or wavelength conversion for phase stabilization or monitoring, particularly because the wavelengths of pump lasers are generally not well-suited for long-distance transmission.
In contrast, the ST, as detailed in the Results section, only requires the pump light to have a coherence time of about $N\mathit{\Delta_{\mathrm{t}}}$, eliminating the need to stabilize or monitor phase differences between distant lasers. 
Therefore, this relaxation of stability requirements is expected to greatly simplify the setup.

\clearpage
\section*{References}
\bibliography{ref}
\bibliographystyle{naturemag2.bst}

\clearpage
\noindent
\section*{Methods}
\subsection*{Calculations of coincidence rate at elementary link}
\begin{figure}[h]
    \centering
    \includegraphics[width=0.5\textwidth]{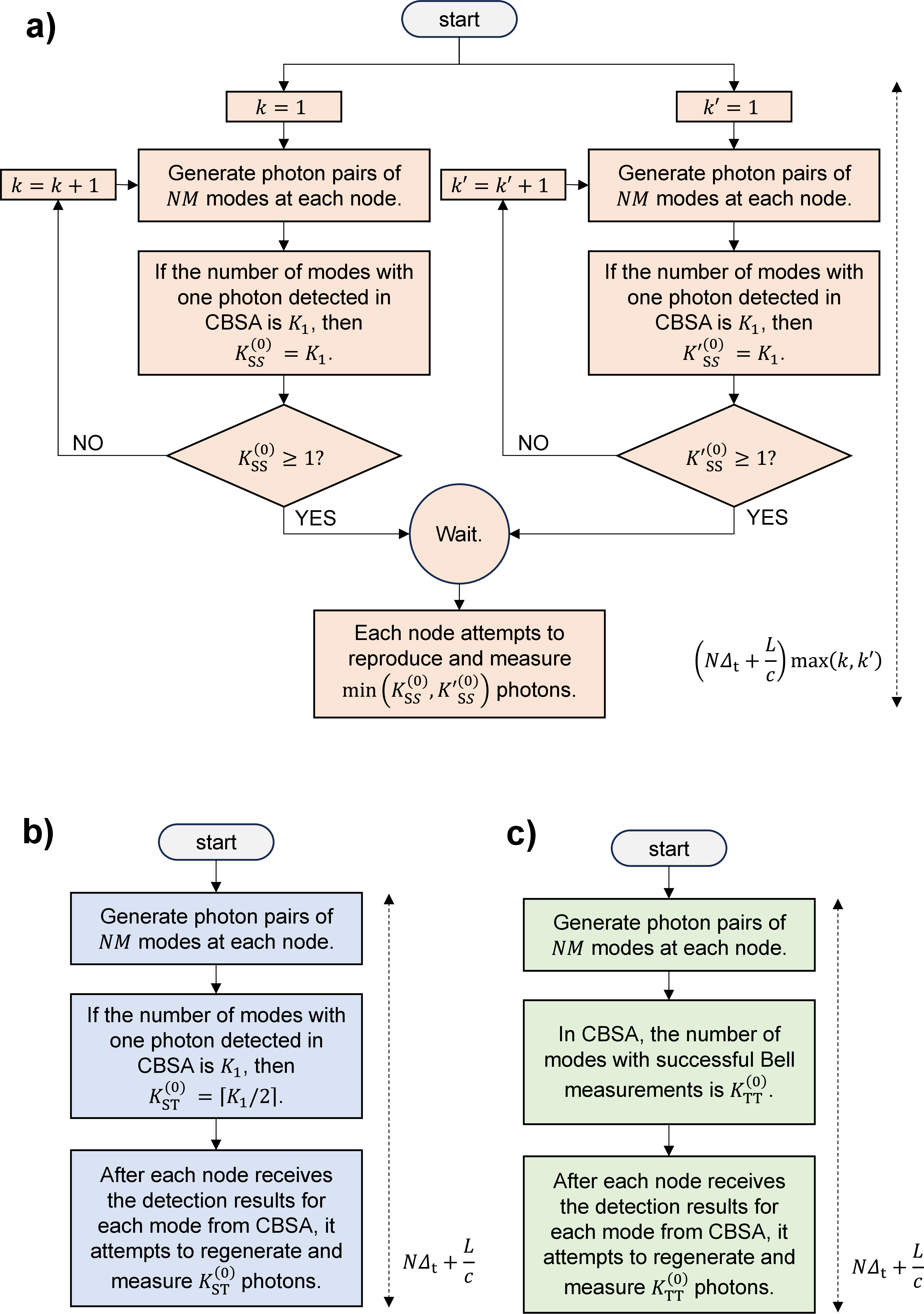}
    \caption{
    {\bf Flowchart of trials in the elementary link.}
    {\bf a}, Flowchart of attempted coinsidence in the SS. 
    $k$ and $k'$ represent the number of iterations until one or more entanglements are heralded in each transmission channel.
    {\bf b} and {\bf c} are flowcharts for attempting cointegration in the ST and TT, respectively.
    }
    \label{chart}
\end{figure}

Fig \ref{chart} illustrates the process of attempting coincidence detection at the elementary link for each scheme.
The duration of one trial, $t^{(0)}$, spans from the start of transmitting the photon in the first temporal mode to the CBSA, until each node receives the result of the photon observation in the $N$-th temporal mode at CBSA, calculated as $N\mathit{\Delta}_\mathrm{t}+L/c$. 

For the SS, the results shown in Fig.~\ref{fig:ele} were generated using a Monte Carlo simulation based on the flow in Fig.~\ref{chart}.
In SS, a method for obtaining coincidences through two transmission paths has been proposed~\cite{Duan2001,Sangouard2011}. 
This method involves heralding single excitation entanglement on both paths, followed by attempts to achieve coincidences.
Let $K^{(0)}_\mathrm{SS}$ and $K'^{(0)}_\mathrm{SS}$ denote the number of heralded entanglements in one trial for each path. 
Coincidence attempts are made $\min \left(K^{(0)}_\mathrm{ST}, K'^{(0)}_\mathrm{ST}\right)$ times after both $K^{(0)}_\mathrm{SS}$ and $K'^{(0)}_\mathrm{SS}$ reach at least 1. 
The probability of a successful coincidence in each attempt is $\eta^2/2$, where $\eta = \eta_\mathrm{QM}\eta_\mathrm{FM}\eta_\mathrm{det}$. 

For the ST and TT, $K_\mathrm{ST}^{(0)}$ and $K_\mathrm{TT}^{(0)}$ represent the number of attempts for coincidence in one trial.
The probability of successful coincidence after heralding differs between ST and TT, as ST may have two excitations on one memory side.
The probabilities are $\eta^2/2$ and $\eta^2$, respectively. 
Thus, the coincidence rates $R_\mathrm{ST}^{(0)}$ and $R_\mathrm{TT}^{(0)}$ for the elementary link in the ST and TT are given by:
\begin{align}
    R_\mathrm{ST}^{(0)} = \frac{\eta^2 E(K_\mathrm{ST}^{(0)})}{2t^{(0)}}, \label{eq:ReleST}
\end{align}
\begin{align}
    R_\mathrm{TT}^{(0)} = \frac{\eta^2 E(K_\mathrm{TT}^{(0)})}{t^{(0)}}. \label{eq:ReleTT}
\end{align}
The results for ST and TT in Fig. \ref{fig:ele} were plotted using Eqs. \eqref{eq:ReleST} and \eqref{eq:ReleTT}.

\section*{Note added.}
During the preparation of this work, a work by Li et al.~\cite{yin2024asynchronous} that employs a similar idea of pairing two single-photon interference events appeared on arXiv.
Our contribution extends this concept by introducing frequency multiplexing and proposing a specific physical system for implementation.

\clearpage
\noindent
{\bf Acknowledgements}\\
We thank Mayuka Ichihara and Koji Nagano for valuable discussions. 
This research was supported by the SECOM Foundation, JSPS KAKENHI (JP20H02652), Deep-Tech Startups Support Program (New Energy and Industrial Technology Development Organization, Japan), JST Moonshot R\&D (JPMJMS226C), and R\&D of ICT Priority Technology Project (JPMI00316).
We also acknowledge the members of the Quantum Internet Task Force, which is a research consortium aiming to realize the quantum internet, for comprehensive and interdisciplinary discussions of the quantum internet.
\\

\noindent
{\bf Author contributions}\\
D.Y. conceived the research. 
D.Y. and T.H. designed the protocol.
D.Y. performed numerical simulation.
All authors discussed the results and commented on the manuscript.
T.H. supervised the project.
\\

\noindent
{\bf Competing interests}\\
All authors have a financial interest in LQUOM Inc.
\\

\noindent
{\bf Materials \& Correspondence}\\
Correspondence and material requests should be addressed to D.Y.

\clearpage

\clearpage

\clearpage

\clearpage
\begin{figure}[h]
    \centering
    \includegraphics[width=0.5\textwidth]{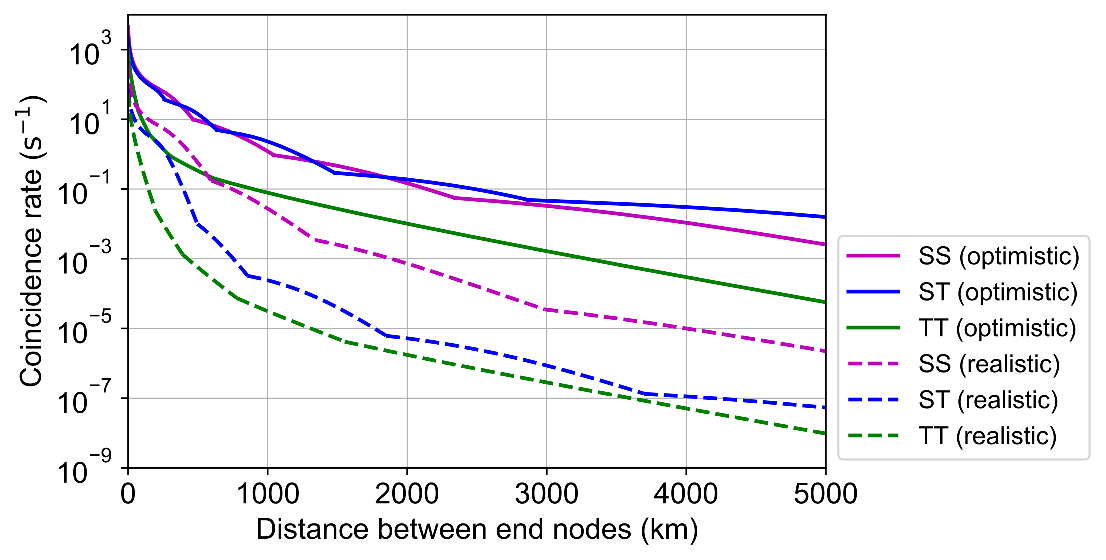}
    \caption{
    {\bf Comparison of each scheme in $2^J$ elementary links.}
    The graph illustrates the coincidence rate at the end nodes of a quantum communication path consisting of $2^j$ elementary links
    The horizontal axis represents the distance between end nodes. 
    The rate at the number of links with the best rate at each end node distance is plotted.
    The solid line depicts results using ideal parameters with values $\eta_{\mathrm{det}} = 0.95$, $\eta_{\mathrm{QM}} = 0.9$, and $\eta_{\mathrm{FM}} = 0.95$.
    The dashed line represents results using realistic parameters with values $\eta_{\mathrm{det}} = 0.9$, $\eta_{\mathrm{QM}} = 0.5$, and $\eta_{\mathrm{FM}} = 0.9$.
    Other parameters were the same for both, $p_{\mathrm{tps}} = 0.01$, $M = 100$, and $N = 30$.
    }
    \label{Figjlink}
\end{figure}

\clearpage

\clearpage
\section*{Supplimentary Information Note 1: Calculation of coincidence rates}
\newcounter{suppequation}
\renewcommand{\theequation}{S\arabic{suppequation}}
\subsection{\label{level2-2} Calculation of coincidence rates for SS schemes with $2^j$ links}
In the context of the elementary link in the SS scheme, after a heralding event occurs in a certain mode, the state between the memories of that mode is denoted as $\rho_\mathrm{SS}^{(0)}=\alpha^{(0)}\rho_\mathrm{1} + (1 - \alpha^{(0)})\rho_\mathrm{0}$.
Assuming negligible dark counts, $\alpha^{(0)}=1$.
 Furthermore, the state between memories after entanglement swapping between $\rho_\mathrm{SS}^{(j-1)}$ is represented as $\rho_\mathrm{SS}^{(j)}=\alpha^{(j)}\rho_\mathrm{1} + (1 - \alpha^{(j)})\rho_\mathrm{0}$.
 Here, $\rho_1$ signifies the desired state with a single excitation entangled state, while $\rho_0$ represents the vacuum state. 
 Additionally, $\alpha^{(0)}$ and $\alpha^{(j)}$  denote the probabilities that $\rho_\mathrm{SS}^{(0)}$ and $\rho_\mathrm{SS}^{(j)}$ are in the desired state, respectively.
 Considering entanglement swapping between  $\rho_\mathrm{SS}^{(j-1)}$ states, the successful event of this swapping between $\rho_\mathrm{SS}^{(j-1)}$  states is characterized by the detection of a single photon in SBSA.
 This event can unfortunately occur when both states are $\rho_{1}$ ,albeit with losses, or when only one of them is $\rho_{1}$.
 These unintended successful events contribute to the vacuum state of $\rho_\mathrm{SS}^{(j)}$.
 The probabilities of successful events of entanglement swapping between $\rho_\mathrm{SS}^{(j-1)}$ states, $p_\mathrm{SS}^{(j)}$, and $\alpha^{(j)}$ are given as
 \stepcounter{suppequation}
 \begin{align}
    p_\mathrm{SS}^{(j)} = \alpha^{(j-1)}\eta\left(1-\left(\frac{\alpha^{(j-1)}\eta}{2}\right)\right), \label{eq:pSS}
\end{align}
\stepcounter{suppequation}
\begin{align}
    \alpha^{(j)} = \frac{\alpha^{(j-1)}}{2-\alpha^{(j-1)}\eta}. \label{eq:alphaSS}
\end{align}
In Equation~\eqref{eq:alphaSS}, the parameter $\alpha^{(j)}$ is determined by dividing the probability of desired successful events in the swapping process, leading to $\rho_1$, by $p_\mathrm{SS}^{(j)}$.
After a pair of $\rho_\mathrm{SS}^{(j)}$ is heralded at the end nodes of a quantum communication channel composed of two parallel $2^j$-elementary links, the probability of successfully attempting a coincidence is given by 
\stepcounter{suppequation}
\begin{align}
    p_\mathrm{ps, SS}^{(j)} = \left(\alpha^{(j)} \eta\right) ^2/2. \label{eq:ppsSS}
\end{align}

By substituting the probabilities $p_\mathrm{SS}^{(j)}$ and $p_\mathrm{ps, SS}^{(j)}$, as presented earlier, into Equation (12) in the main text, the coincidence rate  in the SS scheme is calculated.
Here, the probability $p_\mathrm{SS}^{(0)}$ that at least one entanglement state is heralded on an elementary link is given by
$p_\mathrm{SS}^{(0)} = 1 - (1 - 2p_\mathrm{tps}\eta_\mathrm{det}\eta_\mathrm{att}(L/2))^{NM}$.

\subsection{\label{level2-3}Calculation of coincidence rates for ST schemes with $2^j$ links}
In the ST scheme, the state heralded between the memories of the elementary link is represented as  $\rho_\mathrm{ST}^{(0)}=\alpha_\mathrm{11}^{(0)}\rho_\mathrm{11} + \alpha_\mathrm{20}^{(0)}\rho_\mathrm{20}$.
$\rho_\mathrm{11}$ represents the desired state where both memories are individually excited once, while $\rho_\mathrm{20}$ denotes the undesired state where one memory is excited twice and the other memory is not excited.
Additionally, $\alpha_\mathrm{11}^{(0)}$ and $\alpha_\mathrm{20}^{(0)}$ represent the probabilities of $\rho_\mathrm{ST}^{(0)}$ being in states $\rho_\mathrm{11}$ and $\rho_\mathrm{20}$, respectively.
Next, we consider the swapping of $\rho_\mathrm{ST}^{(0)}$ states. 
To reduce the occurrence of unexpected success events in the swapping process, we employ a BSA based on a type 2 fusion gate \cite{Chen2007, Soudagar2007}. 
In this case, the probability of a successful swapping event is denoted as
\stepcounter{suppequation}
\begin{equation}
\begin{split}
    p_\mathrm{ST}^{(1)}=\eta^2(\alpha_\mathrm{11}^{(0)})^2/2 &+ \eta^2(1-\eta)\alpha_\mathrm{11}^{(0)}\alpha_\mathrm{20}^{(0)}\\
    &+\eta^2(1-\eta)^2(\alpha_\mathrm{20}^{(0)})^2/2. 
    \label{eq:pST1}
\end{split}
\end{equation}
Furthermore, the state between the memories after a successful swap is given by
$\rho_\mathrm{ST}^{(1)}=\alpha_\mathrm{11}^{(1)}\rho_\mathrm{11} + \alpha_\mathrm{1}^{(1)}\rho_\mathrm{1}+\alpha_\mathrm{0}^{(1)}\rho_\mathrm{0}$
, where 
$\alpha_\mathrm{11}^{(1)}=\eta^2(\alpha_\mathrm{11}^{(0)})^2/2p_\mathrm{ST}^{(1)}$, 
$\alpha_\mathrm{1}^{(1)}=\eta^2(1-\eta) \alpha_\mathrm{11}^{(0)} \alpha_\mathrm{20}^{(0)}/p_\mathrm{ST}^{(1)}$
and 
$\alpha_\mathrm{0}^{(1)}=\eta^2(1-\eta)^2(\alpha_\mathrm{20}^{(0)})^2/2p_\mathrm{ST}^{(1)}$.
Moreover,  the state between the memories after entanglement swapping between $\rho_\mathrm{ST}^{(j-1)}$ is represented as $\rho_\mathrm{ST}^{(j)}=\alpha_\mathrm{11}^{(j)}\rho_\mathrm{11} +\alpha_\mathrm{1}^{(j)}\rho_\mathrm{1}+\alpha_\mathrm{0}^{(j)}\rho_\mathrm{0}$.
Here, for cases where $j \geq 2$, the probability of a successful swapping event between $\rho_\mathrm{ST}^{(j-1)}$ is given by
\stepcounter{suppequation}
\begin{equation}
\begin{split}
    p_\mathrm{ST}^{(j)} = 
\eta^2(\alpha_{11}^{(j-1)})^2/2 &+ 
\eta^2\alpha_{11}^{(j-1)}\alpha_{1}^{(j-1)}/2 \\
&+\eta^2(\alpha_{1}^{(j-1)})^2/8, 
    \label{eq:pSTj}
\end{split}
\end{equation}
and the coefficients of $\rho$ are
$\alpha_{11}^{(j)}=\eta^2(\alpha_{11}^{(j-1)})^2/2p_\mathrm{ps, ST}^{(j)}$,
$\alpha_{1}^{(j)}=\eta^2\alpha_{11}^{(j-1)}\alpha_{1}^{(j-1)}/2p_\mathrm{ps, ST}^{(j)}$, and
$\alpha_{0}^{(j)}=\eta^2(\alpha_{1}^{(j-1)})^2/8p_\mathrm{ps, ST}^{(j)}$.
When these calculations are employed, it becomes evident that for $j \geq 2$, the values of $\alpha_{11}^{(j)}$, $\alpha_{1}^{(j)}$, and $\alpha_{0}^{(j)}$ do not depend on $j$.
When attempting coincidence on the state $\rho_\mathrm{ST}^{(j)}$, the success probability is given as
\stepcounter{suppequation}
\begin{align}
    p_\mathrm{ps, ST}^{(j)}=\eta^2(\alpha_{11}^{(j)})^2. 
    \label{eq:ppsSTj}
\end{align}

By substituting the probabilities $p_\mathrm{ST}^{(j)}$ and $p_\mathrm{ps, ST}^{(j)}$, as presented earlier, into Equation (11) in the main text, the coincidence rate in the ST scheme can be calculated.
Here, the probability $p_\mathrm{ST}^{(0)}$ that at least one entanglement state is heralded on an elementary link is given by
$p_\mathrm{ST}^{(0)} = 1 - (1 - 2p_\mathrm{tps}\eta_\mathrm{det}\eta_\mathrm{att}(L/2))^{NM}-
2NMp_\mathrm{tps}\eta_\mathrm{det}\eta_\mathrm{att}(L/2)
(1 - 2p_\mathrm{tps}\eta_\mathrm{det}\eta_\mathrm{att}(L/2))^{NM-1}$.

\subsection{\label{level2-4}Calculation of coincidence rates for TT schemes with $2^j$ links}
In the TT scheme, this study assumes that multiple photon pairs do not occur in a single mode in the each TPS.
Based on this assumption, the state heralded between the memory links of the elementary link is denoted as $\rho^{(0)}_\mathrm{TT}=\rho_\mathrm{11}$. 
Furthermore, when performing entanglement swapping between $\rho^{(0)}_\mathrm{TT}$,
the success probability $p^{(1)}_\mathrm{TT}$ using linear optical elements is $\eta^2/2$.
Subsequently, the state remains $\rho^{(1)}_\mathrm{TT}=\rho_\mathrm{11}$.
Similary, the state after entanglement swapping between $\rho^{(j-1)}_\mathrm{TT}$ is $\rho^{(j)}_\mathrm{TT}=\rho_\mathrm{11}$ with a success probability of
\stepcounter{suppequation}
\begin{align}
    p^{(j)}_\mathrm{TT} = \eta^2/2. 
    \label{eq:pTT}
\end{align}
Finally, when attempting coincidence measurements on the $\rho^{(j)}_\mathrm{TT}$, the success probability is given by
\stepcounter{suppequation}
\begin{align}
    p^{(j)}_\mathrm{ps, TT} = \eta^2. 
    \label{eq:ppsTT}
\end{align}

\end{document}